\documentclass[proceedings, preprint]{rmaa}

\usepackage{paralist}

\usepackage{psfrag,color}

\newcommand{\Ha}{H$\alpha$}  
\newcommand{\Hb}{H$\beta$}

\newcommand{\Msolar}{M$_{\odot}$}                           
 
\SetYear{2008}
\SetConfTitle{A Celebration of Luis Carrasco's 60th Birthday}

\title{On the Early Evolution of Young Starbursts} 

\author{
  Daniel Rosa Gonz\'alez,\altaffilmark{1} 
  Henrique Schmitt\altaffilmark{2}
  Elena Terlevich,\altaffilmark{1}
  and Roberto Terlevich,\altaffilmark{1}
  }

\altaffiltext{1}{
Instituto Nacional de Astrof\'{\i}sica Optica y Electr\'onica.
Luis Enrique Erro No. 1. Tonantzintla, Puebla, C.P. 72840, M\'exico (danrosa@inaoep.mx).
}
\altaffiltext{2}{
Remote Sensing Division, Code 7210, Naval Research Laboratory,
4555 Overlook Avenue, Washington, DC\,20375, USA.}

\shortauthor{Rosa-Gonz\'alez, Schmitt, Terlevich \& Terlevich }
\shorttitle{On the Early Evolution of Young Starbursts}

\listofauthors{D. Rosa-Gonz\'alez, H. Schmitt, E. Terlevich \& R. Terlevich }
\indexauthor{Rosa-Gonz\'alez, D.}
\indexauthor{Schmitt, H.}
\indexauthor{Terlevich, E.}
\indexauthor{Terlevich, R.}

\abstract{
We studied the radio properties of very young massive regions of star 
formation in HII galaxies, with the aim of detecting episodes of recent 
star formation in an early phase of evolution where the first supernovae
start to appear. 
The observed radio spectral energy distribution (SED) covers a behaviour range; 
1) there are  galaxies where the SED
is characterized by a synchrotron-type slope, 2) galaxies with a thermal 
slope, and 3) galaxies with possible free-free absorption at long wavelengths.
The latter SED represents a signature of massive star clusters that are still
well inside the progenitor molecular cloud. 
Based on the comparison of the star formation rates (SFR) determined from
the recombination lines and those determined from the radio emission we find
that SFR(H$\alpha$) is on average five times higher than SFR(1.4~GHz).
These results suggest that the emission of these
galaxies is dominated by a recent and massive star formation event
in which the first supernovae (SN) just started to explode. 
We conclude that the systematic lack of 
synchrotron emission in those systems with the largest equivalent width of \Hb\
can only be explained if those are young starbursts 
 of less than 3.5Myr of age, i.e.~before the first type II SNe  
emerge.
}

\resumen{
En este trabajo estudiamos las propiedades de la emisi\'on en radio de un
conjunto de galaxias HII, con el prop\'osito de detectar regiones de formaci\'on
estelar j\'ovenes, en las cuales las primeras explosiones de supernova est\'an 
teniendo lugar. Las distribuciones espectrales de energ\'\i a observadas tienen
tres comportamientos diferentes: 
1) hay galaxias con una distribuci\'on de energ\'\i a del tipo sincrotr\'on,  
2) galaxias con espectro t\'ermico   
y 3) galaxias que presentan absorci\'on libre-libre a longitudes de onda larga.  
Este \'ultimo caso es un indicador de la presencia de un c\'umulo estelar masivo que se encuentra  
aun dentro de la nube molecular progenitora.  
Comparando las tasas de formaci\'on estelar (SFR) determinadas a partir de lineas de 
recombinaci\'on con las tasas determinadas a partir de las observaciones de
radio encontramos que el SFR(\Ha) es en promedio 5 veces mayor que el SFR(1.4~GHz).
Estos resultados sugieren que la emisi\'on de estas galaxias est\'a dominada
por un episodio de formaci\'on estelar reciente y masivo en el cual las
primeras supernovas estan empezando a explotar. Concluimos que el d\'eficit 
de emisi\'on sincrotr\'on en aquellas galaxias con los mayores anchos
equivalentes de \Hb\,  
solamente puede ser explicado si estos brotes de formaci\'on estelar son extremadamente j\'ovenes
con edades menores que  3.5 millones de a\~nos de forma que aun no han explotado
las primeras supernovas de tipo II.
}

\addkeyword{galaxies: evolution}
\addkeyword{radio continuum: galaxies}

\begin{document}
\maketitle

\section{Introduction}
\label{sec:intro}
Thanks to their proximity and the posibillity of observing them in 
almost all the electromagnetic range, 
nearby galaxies are unique laboratories to study galaxy evolution.   
This contribution focus on the study of early stage of stelar evolution 
when most of the new formed stars remain deep inside the progenitor 
molecular cloud. 
In fact, during the early stages of evolution, very young star
clusters should either lack or have a deficit of synchrotron emission, being dominated
by free-free radio emission. Given that the synchrotron emission observed in
starbursts is due to supernova activity, it is a direct tracer of the end
of the evolution of massive stars (M\,$\ge$ 8\,M$_{\odot}$). In the case of
a coeval star formation burst, the first type II SNe appear
after only 3.5\,Myr and last until the 8\,M$_{\odot}$ stars explode at around
40\,Myr. Consequently, the lack of synchrotron emission in presently starforming regions 
is a good indicator of very young clusters.
HII galaxies are low mass objects whose emission and thus most observables
are dominated by a massive burst of star formation. Their
optical spectrum is identical to that observed in giant HII regions
like 30-Doradus in the LMC. These properties make these galaxies ideal
targets for a study of integrated properties in search for features
related to massive star clusters. Selected sources can then be studied in more detail with
higher resolution observations.
In this sense HII galaxies are considered ``young". In fact they are probably
the youngest stellar systems that can be studied in any detail. The youth 
scenario is supported not only by the strong emission lines from ionized 
gas, but also by their underlying stellar continuum properties, i.e. no stellar
absorptions are detected in those HII galaxies with the strongest emission 
lines (EW(H$\beta) > $150\AA) and only weak hydrogen and helium and some very
weak metal lines are detected on those more evolved HII galaxies
(EW(H$\beta) < $ 50\AA).
The lack of detection of stellar absorptions in the extreme HII galaxies
is due to the fact that the optical spectrum of very young clusters is 
dominated
by the light of massive blue supergiants close to the main sequence 
turn-off.
The spectrum of these stars shows narrow and relatively weak absorptions
of hydrogen and helium plus extremely weak metal lines. In the case
of HII galaxies (and also HII regions) the weak and narrow stellar 
absorptions
are filled with strong emission from the ionized gas of the associated HII 
region
making it impossible to  detect directly their presence.

Here we present the results of a study of integrated radio spectral energy 
distributions (SEDs) of
nascent starburst candidates. We find several galaxies with flat spectra,
typical of thermal emission, or even with inverted spectra, indicating
the presence of heavily embedded star clusters. With the data at hand we
compare the radio and emission line  properties of these
galaxies and propose a qualitative evolutionary model. 

The long term objetives of this project includes not only to find and study 
the youngest bursts in the nearby Universe~\citep{2007ApJ...654..226R} but also  
to understand the transition between the dusty obscured ($z$ between 2 and 3) to 
the optically bright ($z\sim$1) Universe and the potential use of 
HII galaxies as standard candles~\citep{2000MNRAS.311..629M}.

\begin{figure}[!t]
  \includegraphics[width=8cm,height=8.5cm]{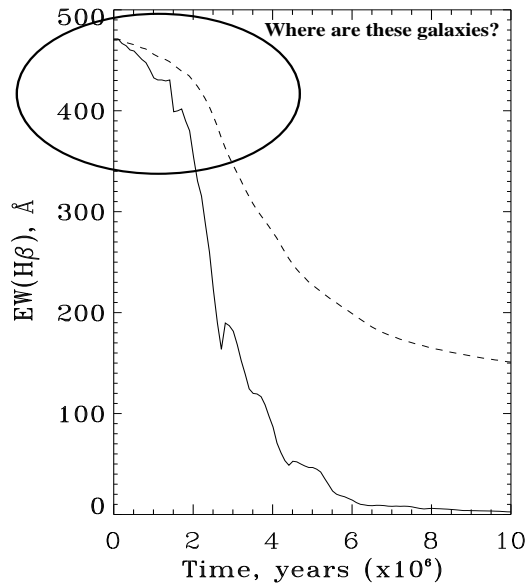}
  \caption{Evolution of the EW(\Hb) for 
an instantaneous burst, solid line, and for continuous star formation, 
dashed line.
}
\label{Fig:EWevol}
\end{figure}

\section{First Evidence of Dealing with Nascent Galaxies}
\label{Sec:Sample}

Our sample selection was based on the expectation that in young starforming galaxies
their most massive  stars did not have enough time to evolve and explode as supernovae. In a
situation like this, these galaxies should have very little synchrotron emission
\citep{2002A&A...392..377B}.
Consequently, their radio luminosities should be smaller than what one
would expect to find based on the SFR measured from other indicators, like
H$\alpha$. 

For the current study we selected a sample from the catalog of HII
galaxies  by~\citet{1991A&AS...91..285T}. 
Based on the reddening corrected
H$\alpha$ fluxes of the galaxies in this catalog, we estimate their Star
Formation Rates (SFR) using the relation given by~\citet{2002MNRAS.332..283R}, 

\begin{equation}\label{eq:SFR_Ha}
\rm SFR [M_\odot year^{-1}]  = 1.1\times 10^{-41}\times L(H\alpha) 
\end{equation}
where the L(\Ha) is given in erg s$^{-1}$.
These SFRs were then converted into radio 1.4~GHz 
fluxes, using the relation given by~\citet{2006ApJ...643..173S}.
Comparing the predicted
1.4~GHz fluxes with values obtained from the NRAO/VLA Sky Survey 
(NVSS) catalog, we selected those
galaxies which were not detected at the NVSS 5$\sigma$ limit (2.5~mJy), or have
an observed flux smaller than the predicted one.
This deficit of 1.4GHz emission can be
quantified in terms of the ratio between the observed 1.4 GHz flux and the
expected flux based on the SFR calculated using the optical emission lines.
For simplicity we denoted this ratio as the $d$-parameter.

The EW(\Hb) has long been used to estimate the age of a stellar burst 
\citep[e.g.][]{1981Ap&SS..80..267D,2001ApJ...553..633L}.
However,  the EW(\Hb) indicator which consists of the relation 
between the continuum flux which depends on the whole 
star formation history of the galaxy and the strength of the emission
line which depends on the recent ($\sim 5\times 10^6$ years) star formation 
activity can  be significantly lower than the age of the current burst.
This problem has been noticed in the analysis of the most extreme  
HII galaxies by~\citet{2004MNRAS.348.1191T}. 
To illustrate this problem, we plot in Figure~\ref{Fig:EWevol} 
the evolution of the EW(\Hb) for the case of an 
instantaneous burst and for  continuous star formation. 
Both models were calculated using the SB99 code
for the case of a Salpeter initial mass function and masses between
0.1 and 100\Msolar. In both modes the EW(\Hb) decreases with time 
but due to the formation of new  massive stars 
in the continuous mode, the EW(\Hb) remains larger as time increases.
These star forming histories  mark the two extrema and the age 
of a  galaxy with a given   EW(\Hb) must lie between the limits
shown in Figure~\ref{Fig:EWevol}.
In  the case of the existence of an old population, the observed  EW(\Hb) 
is reduced even more -- in comparison with the continuous case -- 
due to the integrated light of the galaxy and the absence of massive
stars responsible for the recombination lines. 
Interesting enough is the fact that there are no galaxies known to have 
EW(\Hb) greater than about 350\AA. 

\begin{figure}[!t]
  \includegraphics[width=8cm,height=8.5cm]{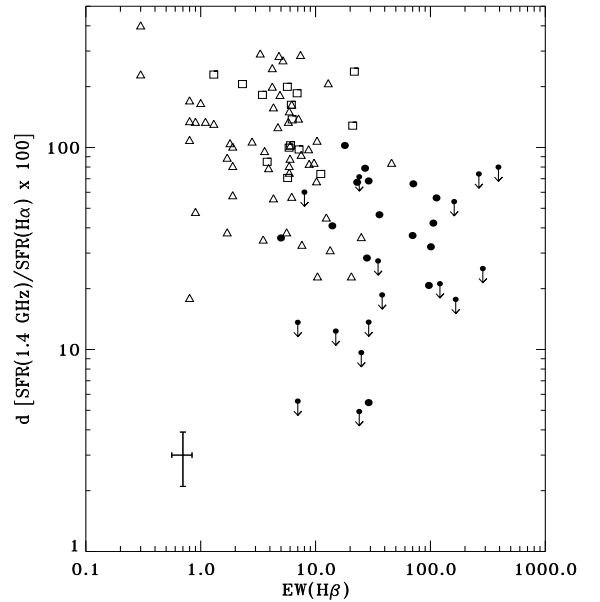}
  \caption{\Hb\, equivalent width against the 
$d$ parameter defined as the ratio between the observed
and the expected  1.4 GHz flux (see text for further details). 
Circles are the sample HII galaxies and 
open symbols represent the control sample 
from Jansen et al. (2000) (triangles) and  from Ho et al.~(1997) (squares).}
\label{SynchrPercent}
\end{figure}

Figure~\ref{SynchrPercent} shows the EW(\Hb) against the $d$-parameter
expressed in percentage. 
Notice that most of our galaxies lie in a region
where the observed flux is less than 50\% of the expected value, showing
that indeed our sample has a deficit of synchrotron radiation.
Five galaxies (Tol\,116-325, MRK\,1315, Tol\,1303-281, Tol\,1304-386 and
Tol\,1358-328) are extreme cases with
the highest EW(\Hb) and just upper limits in the observed 1.4~GHz fluxes.
Notice that, based on the EW(\Hb), these galaxies have ages of less
than 10$^7$ years even in the case of  continuous star formation
(see Figure~\ref{Fig:EWevol}).

In order to investigate how our galaxies compare to normal, quiescent
star forming galaxies in the $d-$parameter~$vs.$\,EW(H$\beta$) diagram,
we obtained radio 1.4~GHz fluxes (from NVSS) for a sample of such galaxies
from~\citet{2000ApJS..126..331J,1997ApJ...487..579H}. 
We can see in Figure~\ref{SynchrPercent} that
the normal galaxies have small EW(H$\beta$) and large $d$ values, occupying
a different region relative to the galaxies in our sample, again consistent
with the interpretation that our galaxies are young and have a
deficit of synchrotron radiation. A particularly
interesting result from this Figure is the fact that a significant portion
of the normal galaxies, those with the smaller EW(H$\beta$), have
$d>$100\%. Although this result seems contradictory, since it implies
a higher radio SFR than what is derived from  H$\alpha$, it can be explained
in view of the lifetime of these two indicators. For example, in the case of
a single burst, the H$\alpha$ emission lasts only 10~Myr, until all the
ionizing photons die. The radio emission on the other hand will last much
longer than that, because of the longer synchrotron emission life time and
the fact that the stars that explode as SN last for $\sim$40~Myr.

\begin{figure*}[!t]
\includegraphics[width=5cm,height=5cm]{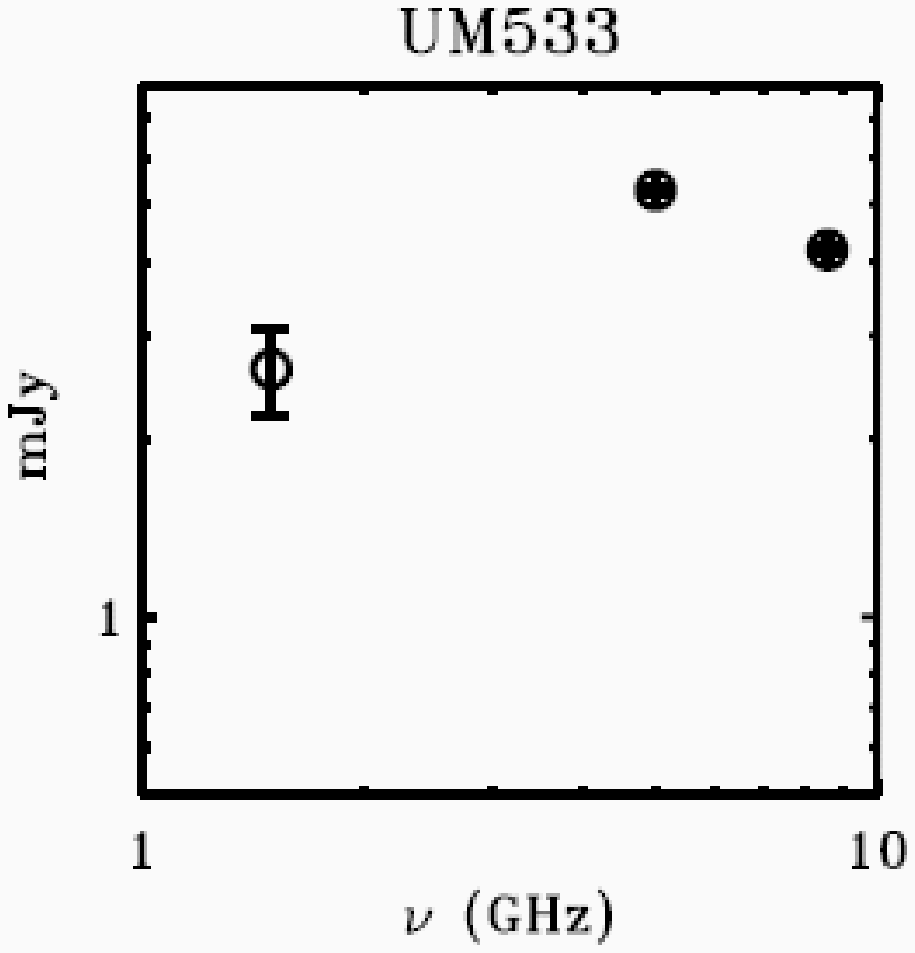}
\includegraphics[width=5cm,height=5cm]{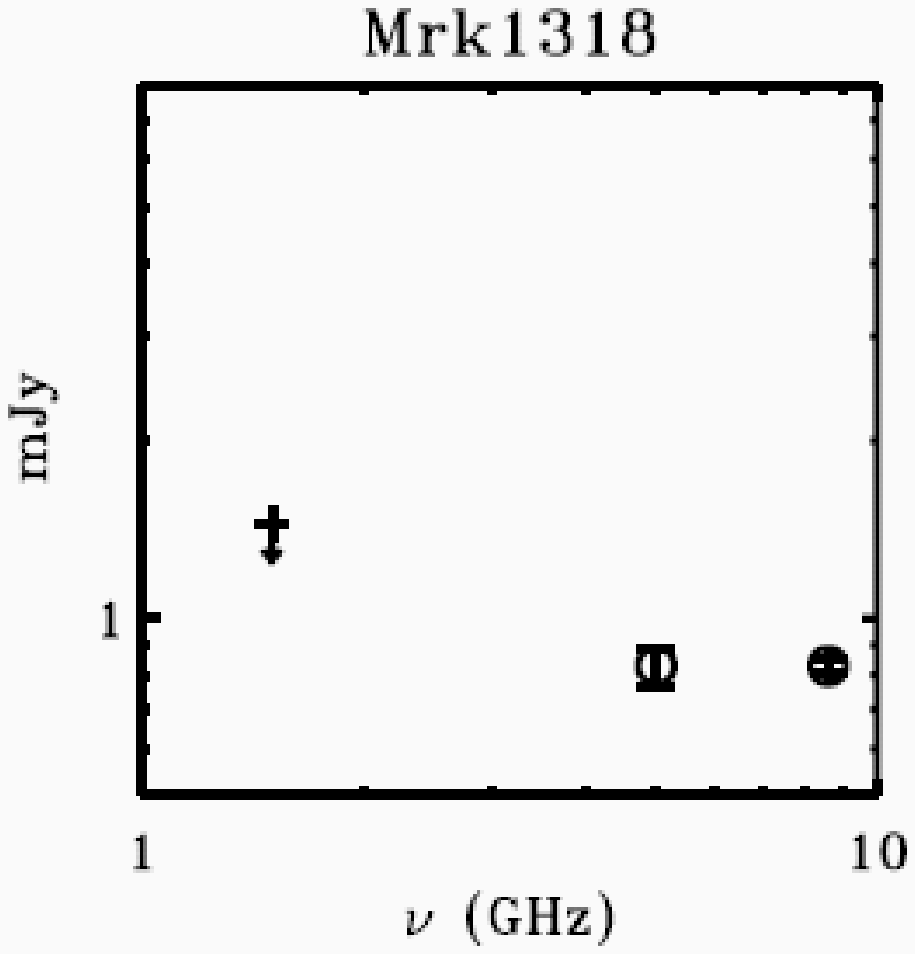}%
\includegraphics[width=5cm,height=5cm]{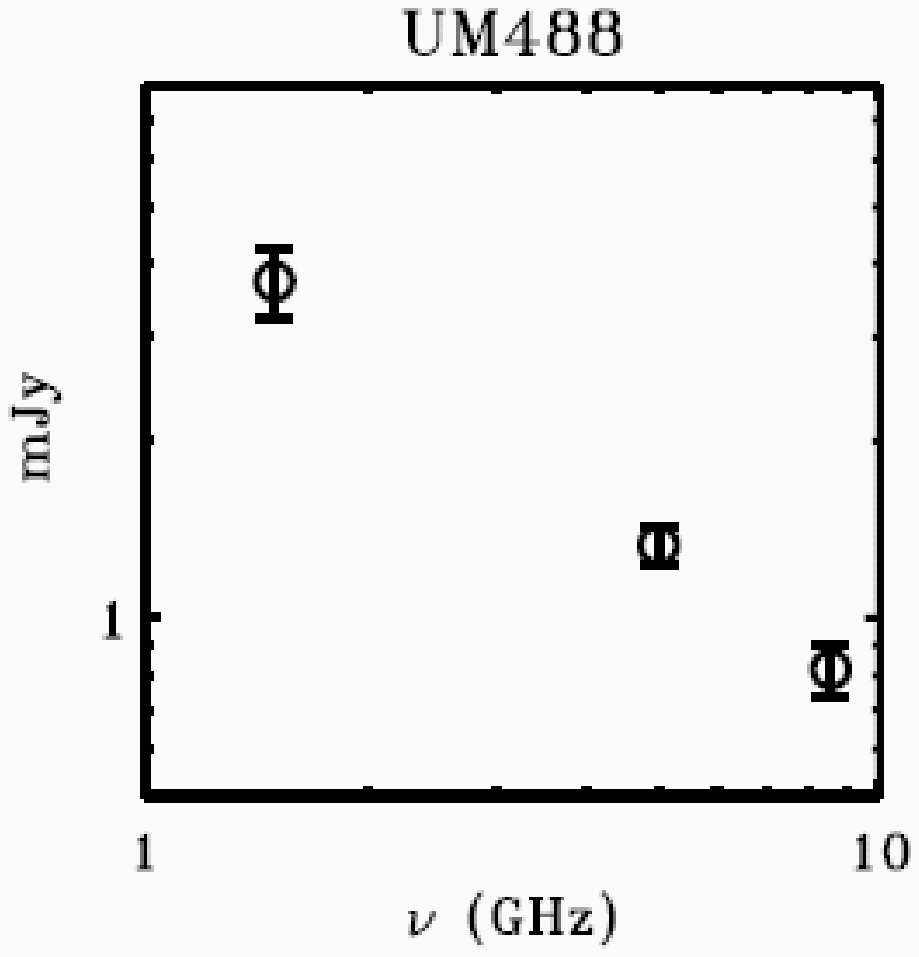}
\caption{Radio SED of three representative galaxies. The SED of UM533 shows
the effect of the free-free absorption which  have reduced significatively the flux at 1.4
GHz. Mrk1318 shows a typical, almost flat, spectrum of a thermal source.
 UM488 shows a  steep spectrum consistent with being dominated
by synchrotron emission.}
  \label{Fig:SEDs}
\end{figure*}

\section{Radio Spectral Energy Distribution }
\label{Sec:Radio}
The galaxies were observed with the VLA at
4.9~GHz (6~cm) and 8.4~GHz (3.5~cm) and we complemented the data set by
including the 1.4~GHz data from the NVSS catalog. 
In Figure~\ref{Fig:SEDs}  we present the radio SEDs
of three selected galaxies. 
Analyzing the spectral slopes of the galaxies detected at 1.4~GHz
we find that they have spectral indices between those expected
for thermal emission ($\alpha=-0.1$), and those characteristic of
a source dominated by synchrotron radiation (typical value is around 
$\alpha=-0.8$, \citeauthor{1992ARA&A..30..575C} 1992.

\begin{figure}[!t]
  \includegraphics[width=8.5cm,height=9cm]{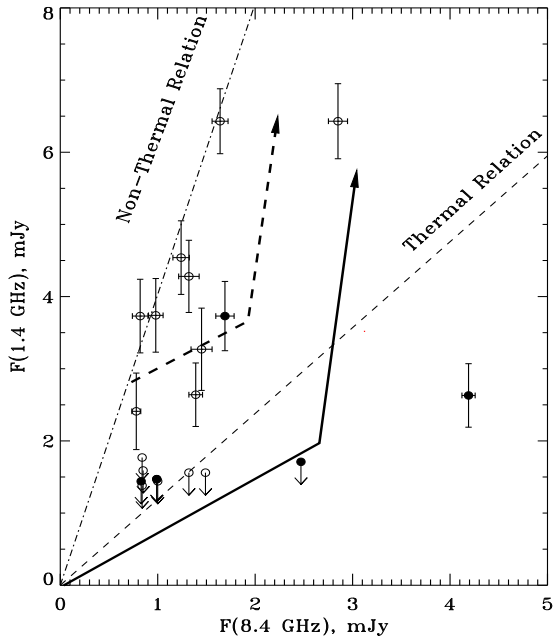}
  \caption{
Distribution of F(1.4~GHz) as a function of F(8.4~GHz) for the selected galaxies. 
The dashed thin line shows the expected
position of  galaxies dominated by thermal emission, while the
dot-dashed line shows where galaxies dominated by non-thermal
emission are located. Superimposed on this plot are solid-thick 
and dashed-thick lines
indicating qualitative evolutionary tracks described in the text.
}
\label{Fig:ThNonTh}
\end{figure}

In Figure~\ref{Fig:ThNonTh} we present the distribution of our
galaxies in the F(1.4~GHz) {\it vs.} F(8.4~GHz) diagram.
This Figure also presents two lines indicating the location of
thermal ($\alpha=-0.1$) and  synchrotron 
($\alpha=-0.8$) emission. In the case of more evolved
star forming galaxies, where we find a mixture of these two
components, the 1.4~GHz emission is dominated by
synchrotron emission, while at 8.4~GHz both synchrotron
and thermal emission produce similar contributions. Most of the
galaxies in our sample lie in the region between these
two lines, indicating that they are young objects still dominated
by thermal emission. 

It is obvious that Figure~\ref{Fig:ThNonTh} represents an oversimplification,
as the history of star formation is much more complicated and cannot be
represented  by either continuous or a single burst or even by a couple of
bursts ~\citep[e.g.][]{2004MNRAS.348.1191T}. Keeping this in mind 
we overplot two qualitative evolutionary tracks to the points in
Figure~\ref{Fig:ThNonTh}. In the first case 
(solid thick line) we assumed 
that the galaxies start with only thermal emission from the OB
associations, which are heavily absorbed at both 1.4~GHz and 8.4~GHz
and can only be detected at the highest frequencies. In this phase the
dense giant HII regions are characterized by optically thick 
free-free emission, which is observed as an inverted spectrum (thermal) radio
source and it is commonly named  ultradense HII region. 
Galaxies for which we only have upper limits
at 1.4~GHz, seem to be  consistent with that description.

Due to the effect of stellar winds, with typical velocities of 
10$^3$ km s$^{-1}$ and mass losses of about 
3 $\times 10^{-5}$\Msolar\ year$^{-1}$,
the absorbing material starts to blow away, the optical depth is reduced 
and  a galaxy moves in this diagram towards the direction of  UM\,533 
(4.2;2.6, in Figure~\ref{Fig:ThNonTh}). 
Once all the absorbing material is blown away, a galaxy moves up in
this diagram, until it reaches the line where the radio emission
is dominated by thermal radiation. 
As the galaxy continues to evolve, SNe
explode producing copious amounts of synchrotron radiation and moving
the galaxies to the top right portion of this diagram where the 
relation between radio and FIR is obtained \citep{2001ApJ...554..803Y}.
In this scenario we expect young galaxies to have $\alpha>-0.1$,
with the younger objects having the larger slopes due to free-free
absorption, evolving towards $\alpha=-0.8$. 

In the second scenario we start with some synchrotron emission due to a
previous generation of supernovae. If a new burst occurs a new generation of
stars is produced, increasing the amount of ionizing photons, and moving the 
galaxy to the right towards the thermal relation (see dashed thick line in 
Figure~\ref{Fig:ThNonTh}). When the next generation of  supernova explodes
the galaxy starts to move up and right to the position of the non-thermal
relation (dot-dashed line). 

\section{Combining  Optical and Radio Measurements}
\label{Sec:OpticalRadio}

In this section we combine the radio and optical data described in 
previous sections in order to test the evolutionary models outlined in
section~\ref{Sec:Radio}.

After an episode of star formation, 
the first SN explosion and consequently the first contribution to the
synchrotron  emission appears at about 3.5$\times 10^6$ years,  
which corresponds to an EW(\Hb) of about 120\,\AA\, for an
instantaneous burst or  310\,\AA\, if
star formation is continuous (Figure~\ref{Fig:EWevol}). 
Therefore objects with high EW(\Hb) and flat slopes are
candidates to be galaxies dominated by very recent starburst
episodes.

\begin{figure}[!t]
\includegraphics[width=8.5cm,height=10cm]{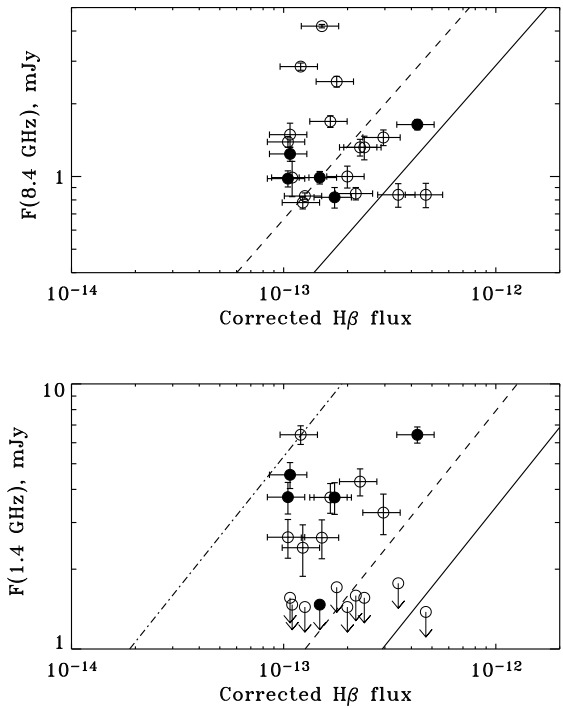}
\caption{
Relation between the extinction corrected \Hb\, flux, the flux at 8.4 GHz 
(top panel) and the flux at 1.4 GHz (bottom panel).  In both panels the solid line is the 
relation between the strength of the \Hb\, line and the corresponding radio flux
for the case of thermal emission coming from a plasma 
with a temperature of $10^4$K. The dashed lines correspond 
to a temperature of 5$\times 10^4$K.
In the bottom panel we draw (dot-dashed-line) the relation given by 
\citet{2001ApJ...554..803Y} which includes synchrotron emission.
In both panels, the solid symbols are galaxies with $\alpha > -0.5$.
Non detections are represented by the arrows (upper limits).
}
\label{Fig:OptRadio}
\end{figure}

Massive stars within a young burst ionize the surrounding media 
producing free electrons and the presence of recombination lines.
The intensity of the \Hb\, emission line is related to the free-free radiation
observed at radio wavelengths  by~\citep{1992ARA&A..30..575C},

\begin{equation}\label{eq:HbRadio}
\rm H\beta\, (erg\, s^{-1}cm^{-2})=  2.8 \times 10^{-11}  (Te/10^4)^{-0.52}  \nu^{0.1}  F_\nu
\end{equation}
where Te is the electron temperature of the plasma,  $\nu$ is the
frequency of the observation (expressed in GHz) and F$_\nu$ is the flux in 
mJy at the given frequency.
For galaxies dominated by synchrotron radiation, the thermal emission is about 10~percent
of the synchrotron radiation at 1.4GHz. However, due to the different 
behaviour and slopes of thermal and non-thermal emission, the thermal emission
at 8.4GHz (3.5cm) could be about 50~percent of the synchrotron one or even more for 
those galaxies that are synchrotron deficient. Figure~\ref{Fig:OptRadio} (top panel) 
shows the extinction corrected \Hb\ flux against the flux at 8.4GHz.
Most of the galaxies have fluxes which are consistent with thermal emission 
from plasmas with temperatures between 10$^4$K (solid line) and  5$ \times
10^4$K (dashed line). 
However some galaxies have an excess of radio emission due to the presence of 
synchrotron emission.

At 20 cm (1.4GHz: bottom panel in Figure~\ref{Fig:OptRadio}) the flux  is
clearly contaminated by the presence of synchrotron emission. 
As in the top panel the solid and dashed lines represent 
the thermal emission based on Equation~\ref{eq:HbRadio}.
All the galaxies detected at 20 cm have fluxes above the thermal limits, 
however the galaxies not detected at 20 cm (the upper limits) 
are close to the thermal region. 

Due to the presence of massive stars in normal and starburst galaxies, there 
is a strong relation between the observed FIR and the flux at 1.4 GHz.
This correlation, which covers several orders of magnitude in
luminosities, is explained by 
the presence of massive stars which heat the dust producing the observed
FIR radiation and, as a consequence of the short life time of these stars, 
they explode as SN producing the synchrotron emission
observed at 1.4 GHz~\citep{2001ApJ...554..803Y}.
Combining this strong correlation with the relation between 
the FIR luminosities and the current SFR from Kennicutt (1998), 
Yun and collaborators proposed a robust relation between 
the luminosities at 1.4~GHz and the SFR,  

\begin{equation}
\rm SFR [M_\odot year^{-1}]  =  5.9 \times 10^{-29} \times L_{1.4 GHz} 
\end{equation}
where the luminosities are expressed in units of  erg\, s$^{-1}$ Hz$^{-1}$.
Notice that all SFR estimates have uncertainties,
because each estimator traces different populations and depends on
several assumptions, e.g.~ the initial mass function,
or on the light attenuation which is a strong function of wavelength.

Comparing the SFR estimated by the 1.4~GHz luminosities with those 
given by the recombination lines 
we find that the optical SFR is on average 4 times higher than 
the SFR given by the radio luminosities.
In the bottom panel of Figure~\ref{Fig:OptRadio} we plot a dot-dashed line 
where the optical SFR is equal to the radio SFR based on the 
Yun et al. (2001) relation. Only three galaxies are close to the 
dot-dashed  line showing that most of the observed starbursts still are 
in the early stages of evolution, when not too many
stars have become SN, and as a result
the observed radio SED is dominated by thermal emission.
The most extreme cases are those galaxies with non detection at 20 cm. 
The cases with clear signature of free-free absorption 
and those dominated exclusively  by thermal emission 
are similar to regions detected in  NGC\,625, and in He2-10.

\section{Discussion and Conclusions}
\label{sec:conclusions}

We combined optical spectroscopy with radio continuum observations to
detect systematically extremely young systems, before the first Type II SNe
explode. The lack of these energetic events produce a deficit in the observed
radio fluxes that was quantified introducing the $d$-parameter. 

To fully explain the observed trends, a highly synchronized 
star formation is required, i.e. the massive star forming phase
should be shorter than $\sim$3-4Myr in order to prevent the appearance of 
the first supernovae
associated to the more massive stars and the consequent synchrotron emission.
This also suggests that there has been no large star forming event in the
previous $10^8$yr.
We just started to put the observational basis to understand the early
evolution of massive stellar clusters and we 
proposed a simplified evolutionary scenario in line with previous theoretical studies. 

The results summarized above have important implications 
for the  understanding of the early stages of evolution of 
low mass galaxies as the ones described in this study and of young massive star
clusters. 

The extreme radio SEDs of these young systems must be included in the
estimation of photometric redshifts of distant obscured galaxies that will be 
discover with the near future facilities (e.g. ALMA and LMT/GTM).
In fact, high redshift (sub)mm-galaxies with extremely weak optical counterparts could
be the analogous of the galaxies discussed in this talk (with a stellar mass $\sim$100 times higher !!!).

Because the high energy (2 -- 10 keV) emission of star-forming
galaxies is mainly due to HMXBs, for a given SFR, the lack of X-ray emission
could be a sign of a young system in which the population of compact objects
is not fully developed~\citep{2007MNRAS.379..357R}.

An extensive version of this work that includes
details of the VLA observations, together 
with further discussions which incorporate among others the analysis 
of the FIR emission and the $q$-parameter has been presented in 
\citet{2007ApJ...654..226R}.

\end{document}